\documentclass{emulateapj}
\usepackage{graphicx}
\usepackage[flushleft]{threeparttable}
\usepackage[usenames,dvipsnames,svgnames,table]{xcolor}
\usepackage{hyperref}
\definecolor{darkblue}{rgb}{0.0,0.0,0.3}
\hypersetup{colorlinks,breaklinks,
            linkcolor=darkblue,urlcolor=darkblue,
            anchorcolor=darkblue,citecolor=darkblue}
\usepackage{amsmath,amssymb}
\usepackage{amsmath}

\begin{document}
\def\actaa{Acta Astronomica}
\def\etal{et al.\ \rm}
\def\ba{\begin{eqnarray}}
\def\ea{\end{eqnarray}}
\def\etal{et al.\ \rm}
\def\Fdw{F_{\rm dw}}
\def\Tex{T_{\rm ex}}
\def\Fdis{F_{\rm dw,dis}}
\def\Fnu{F_\nu}
\def\FJ{F_{J}}
\def\FJE{F_{J,{\rm Edd}}}

\title{Protoplanetary Disk Heating and Evolution Driven by the Spiral Density Waves}

\author{Roman R. Rafikov\altaffilmark{1}}
\altaffiltext{1}{Institute for Advanced Study, Einstein Drive, Princeton, NJ 08540; 
rrr@ias.edu}


\begin{abstract}
High-resolution imaging of some protoplanetary disks in scattered light reveals presence of the global spiral arms of significant amplitude, likely excited by massive planets or stellar companions. Assuming that these arms are density waves, evolving into spiral shocks, we assess their effect on the thermodynamics, accretion, and global evolution of the disk. We derive analytical expressions for the direct (irreversible) heating, angular momentum transport, and mass accretion rate induced by the disk shocks of arbitrary strength. We find these processes to be very sensitive to the shock amplitude. Focusing on the waves of moderate strength (density jump at the shock $\Delta\Sigma/\Sigma\sim 1$) we show the associated disk heating to be negligible (contributing at $\sim 1\%$ level to the energy budget) in passive, irradiated protoplanetary disks on $\sim 100$ AU scales, but becoming important within several AU from the star. At the same time, shock heating can be a significant (or even dominant) energy source in disks of cataclysmic variables, stellar X-ray binaries, and supermassive black hole binaries, heated mainly by viscous dissipation. Mass accretion induced by the global spiral shocks is comparable to (or exceeds) the mass inflow due to viscous stresses. Protoplanetary disks featuring prominent global spirals must be evolving rapidly, in $\lesssim 0.5$ Myr at $\sim 100$ AU. A direct upper limit on the disk evolution timescale can be established via the measurement of the gravitational torque due to the spiral arms from the imaging data. Our findings suggest that, regardless of their origin, global spiral waves must be important agents of the protoplanetary disk evolution. They may serve as an effective mechanism of disk dispersal and could be related to the transitional disk phenomenon. 
\end{abstract}



\section{Introduction.}  
\label{sect:intro}

Recent high-resolution observations of protoplanetary disks using adaptive optics in the optical and near-IR bands have revealed a richness of the non-axisymmetric structures, conspicuous in scattered light images of these systems. Direct imaging in these bands is sensitive to the starlight scattered by the disk, thus such observations probe predominantly the structure of the disk surface.

Some of the most remarkable non-axisymmetric structures detected in scattered light are global spiral arms seen on scales of tens to hundreds of AU \citep{Muto,Grady,Boccaletti,Wagner}. They have also been seen in {\it ALMA} observations of the CO line emission from the disk of HD 142527 \citep{ALMA_spirals}. Spiral arms typically extend over a significant range of radii (at least tens of AU) and exhibit rather open morphology, with pitch angles $\zeta\sim 15^\circ-30^\circ$ \citep{Garufi,Wagner}. Spirals often feature two arms and have surface brightness contrasts ranging from tens of per cent to a factor of a few \citep{Muto,Grady}. However, connecting these fluctuations of the disk brightness to perturbations of the surface density is not trivial: the former probe the corrugations of the scattering disk surface, which are more sensitive to the vertical motions in the disk than to the inhomogeneities of the surface density \citep{Juhasz,Dong_spiral}. Presence of the spiral structure has also been inferred in other types of accretion disks (see \S \ref{sect:other}).

Since the first discoveries of spirals in protoplanetary disks it has been hypothesized that these global structures are density waves excited by massive perturbers, most likely giant planets \citep{Muto,Grady,Benisty}. In some cases stellar companions were shown to be the drivers of these spirals \citep{Dong_star}. Early attempts to match the detailed shapes of the arms using linear theory of global spiral waves  \citep{Rafikov02} met with considerable problems --- they predicted disks to be too hot \citep{Grady,Benisty} and sometimes required rather contrived orbital configurations of planets driving the spiral waves \citep{ALMA_spirals,Benisty}. However, recent 3D simulations of disks perturbed by massive planets ($\gtrsim 5M_J$) and stellar companions have shown much better agreement with observations in terms of the pitch angle of the spiral arms \citep{Dong_spiral,Dong_star}. This agreement has been linked \citep{Zhu} to the importance of the {\it nonlinear} effects for the density wave propagation in disks \citep{GR01,Rafikov02}. These studies have also demonstrated that the observed spiral structures are likely produced by massive perturbers orbiting {\it outside} the spiral arms. 

Apart from revealing possible presence of planets, another interesting aspect of the observed spiral waves is their effect on the disk itself. Density waves carry momentum and energy, which is ultimately shared with the disk. This exchange affects the global properties of the disk and leads to its evolution, as shown by \citet{GR01} and \citet{Rafikov02}. Understanding these effects is the subject of this work. 

Here we focus primarily on the effect of global density waves on the disk thermodynamics and evolution. As the observed waves have rather large amplitudes they are expected to rapidly evolve into shocks due to the nonlinear wave steepening. Dissipation at the shocks provides the means of the momentum and energy exchange between the wave and the disk, which we explore in this work.  

Throughout this study we will assume spiral waves to have already evolved into shocks. Our calculations are three-dimensional, however, later we will often treat the disk as a two-dimensional structure, which does not affect the generality of our conclusions. Our consideration applies to spirals of arbitrary origin (driven by planets, stellar companions, gravitational instability, etc.), with the only important input parameters being the shock strength (chracterized by the pressure and density jumps across the shock) and spiral pattern speed $\Omega_P$. If the spiral is driven by a massive perturber on a circular orbit then $\Omega_P$ should be equal to the angular frequency of that orbit. We will implicitly assume that the perturber is {\it external}, i.e. it orbits outside the observed spiral arms, although this limitation is not essential to our consideration. 

In making numerical estimates we will use disk surface density profile typical for the minimum mass Solar nebula
\ba
\Sigma(r)=10~\mbox{g cm}^{-2}r_{50}^{-3/2},
\label{eq:MMSN}
\ea
where $r_{50}\equiv r/(50$AU) is the (scaled) distance from the star. Such disk contains about $0.05M_\odot$ within 100 AU. 

Brightness asymmetries in scattered light images of disks may also be caused by other mechanisms, e.g. by warps or bending waves \citep{Marino}. Here we do not consider this possibility and assume instead that perturbation is driven primarily by the in-plane fluid motions. This situation is naturally expected if the wave is excited by the perturber, which is co-planar with the disk.

Our work is structured as follows. In \S \ref{sect:heating} we compute the rate of irreversible disk heating by the shock, while in \S \ref{sect:evolution} we explore angular momentum transport and calculate the shock-driven mass accretion rate through the disk. In \S \ref{sect:role_heat} we explore the effect of shock heating on the disk thermodynamics, while in \S \ref{sect:accr_shock}  and \S  \ref{sect:role_evol} we consider the shock-driven mass accretion and global evolution of the disk. Implications for disk dispersal, origin of the transitional disks and planet formation are discussed in \S \ref{sect:impl}. In \S \ref{sect:grav_torque} we evaluate the gravitational torque due to the self-gravity of the non-axisymmetric spiral and then discuss the implications of our results for other types of accretion disks in \S \ref{sect:other}.


\section{Shock-driven disk heating.}  
\label{sect:heating}


We start by computing the rate at which a shock wave injects energy into the fluid that passes through it. Except for \S \ref{sect:heat_rate} our calculations are fully general and do not use the fact that the shock propagates in a differentially rotating disk.  

After passing through the shock gas with the pre-shock density $\rho_0$ and temperature $T_0$ gets compressed to the post-shock density $\rho$ and heated to temperature $T$, given by \citep{LL}
\ba
\frac{\rho}{\rho_0}=\frac{\gamma-1+(\gamma+1)\Pi}{\gamma+1+(\gamma-1)\Pi},~~~~\frac{T}{T_0}=\Pi\frac{\rho_0}{\rho},
\label{eq:Trho}
\ea
where $\Pi\equiv p/p_0$ is the ratio of pressures after ($p$) and prior ($p_0$) to passing the shock. As a result, the change of the  internal energy of gas across the shock per unit mass, $d\Delta U/dm=c_V(T-T_0)$ (where $c_V=k_B/\mu(\gamma-1)$ is the specific heat capacity), is 
\ba
\frac{d\Delta U}{dm} &=& \frac{k_B(T-T_0)}{(\gamma-1)\mu}=\frac{k_BT_0}{\mu}\psi_U(\Pi),
\label{eq:dE}\\
\psi_U(\Pi) &\equiv & \frac{\Pi^2-1}{\gamma-1+(\gamma+1)\Pi}.
\label{eq:psiE}
\ea

Shock heating is accomplished via two processes: (1) adiabatic compression at the shock as the density goes from $\rho_0$ to $\rho$ and (2) irreversible heating due to the microscopic processes at the shock front. The addition of entropy to the disk that serves as the irreversible heating source occurs only during the second process.

\begin{figure*}
\centering
\includegraphics[width=0.9\textwidth]{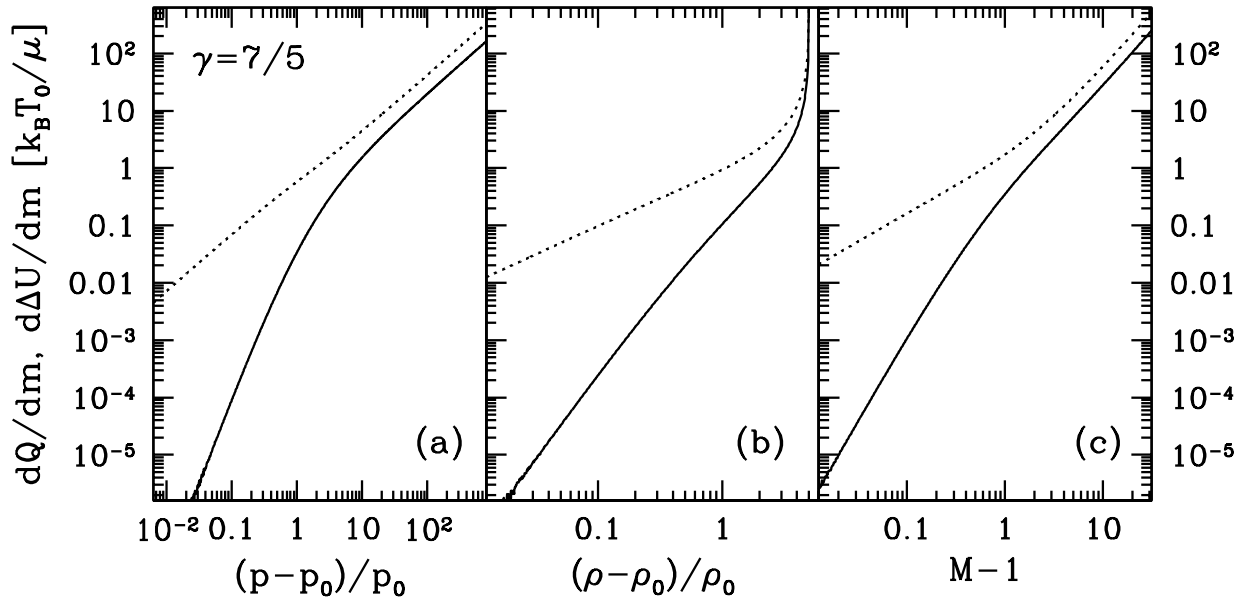}
\caption{Irreversible heating $dQ/dm$ ({\it solid}) and change of the internal energy across the shock $d\Delta U/dm$ ({\it dotted}) per unit mass, in units of $k_BT_0/\mu$ (i.e. $\psi_Q$ and $\psi_U$, respectively). Both are shown as functions of the pressure jump ({\it left}), density jump ({\it center}), and the pre-shock Mach number ({\it right}). Calculation is done for $\gamma=7/5$, appropriate for cool protoplanetary disks.
\label{fig:7_5}}
\end{figure*}

Adiabatic heating is irrelevant as the heat source for the disk since shortly after passing the spiral shock gas density returns to its unperturbed value $\rho_0$ (to be shocked again multiple times in the future), as a result of 2D (or 3D) fluid motions. This is inevitable because of the {\it periodicity} of gas motion in the disk. Such decompression back to the original density adiabatically removes the corresponding share of the temperature increase following each shock crossing. For this reason, in the following we will define shock heating per unit mass $dQ/dm$ to correspond only to the {\it irreversible} part of the temperature increase at the shock. In other words, $dQ/dm=c_V(T_s-T_0)$, where $T_s$ is the gas temperature behind the shock in the absence of adiabatic heating. 

To find $T_s$ one simply needs to find the disk temperature that the shocked gas would have when it is adiabatically expanded to bring its density from $\rho$ back to the original density $\rho_0$. This allows us to find
\ba
T_s=T\left(\frac{\rho_0}{\rho}\right)^{\gamma-1}=T_0~\Pi\left[\frac{\gamma+1+(\gamma-1)\Pi}{\gamma-1+(\gamma+1)\Pi}\right]^{\gamma}.
\label{eq:T_s}
\ea  
Another way to express $T_s$ is via the entropy jump at the shock $\Delta S$:
\ba
T_s=T_0\exp\left(\Delta S/c_V\right).
\label{eq:TsdS}
\ea
This formula uses the fact that in the end of adiabatic decompression gas density goes back to $\rho_0$, while the entropy is still higher than in the pre-shocked gas by $\Delta S$. Expressing $\Delta S=c_V\ln[\Pi(\rho_0/\rho)^\gamma]$ via $\Pi$ using equations (\ref{eq:Trho}) and plugging it into equation (\ref{eq:TsdS}) one again arrives at the result (\ref{eq:T_s}).

Having found $T_s$, we are now in position to determine the irreversible heating produced at the shock, per unit mass crossing the shock: 
\ba
\frac{dQ}{dm} &=& \frac{k_B(T_s-T_0)}{(\gamma-1)\mu}= \frac{k_BT_0}{\mu}\psi_Q(\Pi),
\label{eq:dQ}
\ea
where we have defined
\ba
\psi_Q(\Pi) &\equiv & \frac{1}{(\gamma-1)}\left[\Pi\left[\frac{\gamma+1+(\gamma-1)\Pi}{\gamma-1+(\gamma+1)\Pi}\right]^{\gamma}-1\right].
\label{eq:psiQ}
\ea  
These equations represent one of the key results of our work.

Another convenient way to express irreversible heating is via the density contrast across the shock $\rho/\rho_0$, which can be done using equation (\ref{eq:Trho}):
\ba
\psi_Q = \frac{1}{(\gamma-1)}\left[\left(\frac{\rho_0}{\rho}\right)^{\gamma}
\frac{(\gamma+1)(\rho/\rho_0)-(\gamma-1)}{(\gamma+1)-(\gamma-1)(\rho/\rho_0)}-1\right].
\label{eq:psiQ_rho}
\ea 

In the case of isothermal equation of state one finds, taking the limit $\gamma\to 1$, that
\ba
\psi_Q &=& \frac{\epsilon(2+\epsilon)-2(1+\epsilon)\ln(1+\epsilon)}{2(1+\epsilon)},
\label{eq:isoth}
\ea
where $\epsilon \equiv (\rho-\rho_0)/\rho_0=(p-p_0)/p_0$. This expression is valid for the arbitrary value of $\epsilon$, and agrees with similar result found in \citet{Belyaev2}.

\begin{figure*}
\centering
\includegraphics[width=0.9\textwidth]{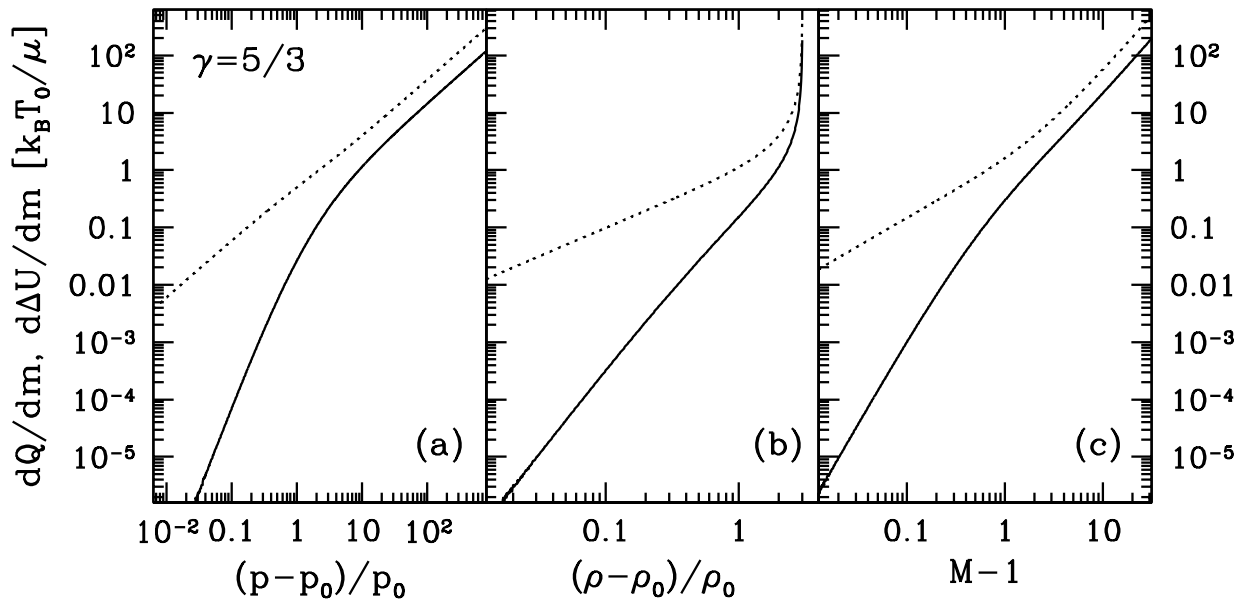}
\caption{Same as Figure \ref{fig:7_5} but for $\gamma=5/3$, as expected for hot accretion disks studied in \S \ref{sect:other}.
\label{fig:5_3}}
\end{figure*}

In Figures \ref{fig:7_5} and \ref{fig:5_3} we plot both $dQ/dm$ and $d\Delta U/dm$ for $\gamma=7/5$ (typical for protoplanetary disks) and $5/3$ (expected for hotter disks), respectively, as functions of the pressure jump $\Delta p/p_0\equiv (p-p_0)/p_0$, density jump $\Delta\rho/\rho_0\equiv (\rho-\rho_0)/\rho_0$, and the pre-shock Mach number $M$ \citep{LL}
\ba
M^2=1+\frac{\gamma+1}{2\gamma}\frac{p-p_0}{p_0}.
\label{eq:Mach}
\ea
One can clearly see the difference in $dQ/dm$ and $d\Delta U/dm$ behaviors, which we discuss in more details next. Note that $\psi_Q\lesssim 1$ even for shocks of moderate strength with $\Delta \rho/\rho_0\sim 1$. For example, for $\gamma=7/5$ one has $\psi_Q\approx 0.1$ for $\Delta\rho/\rho_0=1$, while already for $\Delta\rho/\rho_0=0.3$ one finds $\psi_Q\approx 0.005$.


\subsection{Weak shocks.}  
\label{sect:weak}


The distinction between the irreversible heating $dQ/dm$ and the full change of internal energy $d\Delta U/dm$ is most pronounced in the case of {\it weak} shocks, when $\Delta p/p_0\ll 1$. In this limit equations (\ref{eq:psiE}) and (\ref{eq:psiQ}) reduce to 
\ba
\psi_U\approx \gamma^{-1}\frac{\Delta p}{p_0}\approx \frac{\Delta\rho}{\rho_0},
\label{eq:Eweak}
\ea
and 
\ba
\psi_Q &\approx &\frac{\gamma+1}{12\gamma^2}\left(\frac{\Delta p}{p_0}\right)^3
\approx \frac{\gamma(\gamma+1)}{12}\left(\frac{\Delta\rho}{\rho_0}\right)^3
\nonumber\\
&\approx & \frac{2\gamma}{3(\gamma+1)^2}\left(M^2-1\right)^3,
\label{eq:Qweak}
\ea
where we used the relation $\Delta p/p_0\approx \gamma \Delta\rho/\rho_0$ valid for weak shocks. Equation (\ref{eq:Qweak}) represents a well known result \citep{LL} that the irreversible heat release and entropy production by the weak shock scale as the third power of the pressure or density perturbation, explaining the aforementioned smallness of $\psi_Q$ even for $\Delta\rho/\rho\lesssim 1$. Equation (\ref{eq:Qweak}) coincides to leading order with the similar result obtained by \citet{Larson_conf,Larson}.

Comparison of equations (\ref{eq:Eweak}) and (\ref{eq:Qweak}) makes it obvious that in weak shocks $dQ/dm\ll d\Delta U/dm$ and irreversible heat generation contributes negligibly to the internal energy increase. After the shocked gas returns to its initial density $\rho_0$, the amount of heat $dQ/dm$ injected by the weak shock into the disk is going to be very small.

\subsection{Strong shocks.}  
\label{sect:strong}


In the opposite limit of {\it strong} shocks, $\Pi=p/p_0\gg 1$, we find 
\ba
\psi_U\approx\frac{\Pi}{\gamma+1},~~~\psi_Q\approx\left(\frac{\gamma-1}{\gamma+1}\right)^{\gamma}\frac{\Pi}{\gamma-1}.
\label{eq:psi_strong}
\ea
Thus, both $d\Delta U/dm$ and $dQ/dm$ scale linearly with the pressure ratio across the shock $p/p_0$, and the latter constitutes a fixed fraction $[(\gamma-1)/(\gamma+1)]^{\gamma-1}$ of the former. This relative contribution of the irreversible heating is $\approx 0.488$ for $\gamma=7/5$ and $\approx 0.397$ for $\gamma=5/3$.


\subsection{Disk heating rate.}  
\label{sect:heat_rate}


Equations (\ref{eq:dQ})-(\ref{eq:psiQ_rho}) give the rate of irreversible energy injection per unit mass crossing the shock. At the same time, for practical purposes it is important to know the rate of energy release per unit time and radius in the disk, $d \dot Q/d r$. To compute this quantity we must calculate the amount of mass $dm$ within radial interval $dr$ that passes through the shock per unit of time $dt$. For a spiral wave with pattern speed $\Omega_P$ and $m$-fold azimuthal symmetry ($m$ arms) this mass is $dm=m|\Omega_P-\Omega(r)|r\Sigma dt dr$, where $\Sigma$ is the surface density of the disk and $\Omega$ is its angular frequency. Note that $dm$ is independent of the pitch angle of the spiral shock, as long as the radial velocity of the disk fluid is small.

Combining this with equation (\ref{eq:dQ}) we find
\ba
\frac{d\dot Q}{d r} &=& m|\Omega_P-\Omega(r)|r\Sigma\frac{dQ}{dm}
\nonumber\\
&=& m|\Omega_P-\Omega(r)|r~\Sigma c_0^2~\psi_Q(\Pi),
\label{eq:heat_per_r}
\ea
where $c_0\equiv k_BT_0/\mu$ is the pre-shock {\it isothermal} sound speed. Note that, again, the pitch angle of the shock does not enter this expression, only its pattern speed $\Omega_P$ and the pressure contrast $\Pi$.

The spiral shock is a non-axisymmetric structure. However, on timescales much longer than $|\Omega_P-\Omega|^{-1}$ any fluid element in the disk gets shocked multiple times, and in the {\it time-averaged} sence it is possible to define the heating rate per unit surface area $dS$ (and unit of time) $d\dot Q/d S =(2\pi r)^{-1}d\dot Q/d r$. Using equation (\ref{eq:heat_per_r}) one finds
\ba
\frac{d\dot Q}{d S}=\frac{m}{2\pi}|\Omega_P-\Omega(r)|~\Sigma c_0^2~\psi_Q(\Pi).
\label{eq:heat_per_S}
\ea
We will discuss the applications of this expression to protoplanetary disks in \S \ref{sect:role_heat} and to other types of accretion disks in \S \ref{sect:other}.

Equation (\ref{eq:heat_per_r}) describes the {\it direct} injection of entropy and energy into the disk by the shock, which is not the same as $\Delta U$, as we discussed earlier. However, the post-shock temperature increase may still cause additional {\it indirect} heating or cooling of the disk. Indeed, protoplanetary disks are illuminated by their central stars, which provide their main energy source (see \S \ref{sect:role_heat}). Post-shock temperature jump drives vigorous vertical motions behind the strong shock \citep{Boley,Zhu} causing substantial corrugation of the scattering surface of the disk \citep{Dong_spiral}. This leads to interception of more starlight by the disk surface locally, resulting in enhanced {\it indirect heating} by irradiation at this location. However, global energy balance of the disk is unlikely to be affected by this process, as long as the density wave does not change the full solid angle subtended by the disk from the vantage point of the star. 

Moreover, the post-shock temperature rise should extend to the disk photosphere. This temporarily (until temperature goes back to its unperturbed value) increases radiative losses from the disk surface resulting in additional {\it indirect cooling} of the disk by the post-shock temperature jump. 

These indirect contributions to the thermal energy budget of the disk depend on the full $\Delta U$. However, given their sensitivity to the details of the calculation (disk structure, geometry of illumination, and so on) we do not consider them in this work and focus only on the direct irreversible heating of the disk.


\section{Shock-driven angular momentum and mass transport.}  
\label{sect:evolution}


Irreversible dissipation at the shock results in the change of the angular momentum of the fluid that passes through the shock. This leads to a radial mass flux through the disk, ultimately resulting in its evolution. Once again, even though the disk shock itself is a non-axisymmetric structure, on timescales much longer than the local orbital timescale $\Omega^{-1}$ one can characterize disk evolution via the azimuthally averaged mass accretion rate $\dot M\equiv \int_0^{2\pi}\Sigma v_r d\varphi$, where $\Sigma$ is the disk surface density and $v_r$ is the radial velocity of the fluid. 

\citet{Belyaev2} derived an expression for $\dot M(r)$ driven by the weak shocks in the disk, see their \footnote{Note that their Equation (B8) for the accretion timescale $t_{\rm acc}$ has a typo --- an extra factor $\varpi^{-1}$, which should not be there.} Appendix B. Here we generalize calculation of $\dot M$ to the case of an arbitrary shock strength. The main steps are the same as in \citet{Belyaev2}, but for completeness we briefly outline them here. 

Mass accretion rate is related to $d F_J/d r$ --- the rate of angular momentum {\it deposition} into the disk fluid per unit radius. As shown by \citet{lynden-bell_1974},
\ba
\dot M(r)=\left(\frac{dl}{dr}\right)^{-1}\frac{d F_J}{d r},
\label{eq:Mdot}
\ea
where $l=\Omega r^2$ is the specific angular momentum of the disk fluid. Injection of angular momentum can be effected in a variety of ways, including viscosity (molecular or anomalous, see \S \ref{sect:accr_shock} below), but here we focus on the angular momentum transfer due to the spiral shocks.

To find $d F_J/d r$ we use the relation \citep{GR01}
\ba
\frac{d\dot Q}{d r} =[\Omega(r)-\Omega_P]\frac{d F_J}{d r}, 
\label{eq:heat_J}
\ea
which follows from the fact that external potential does work $\Omega_P\Delta J$ to inject angular momentum $\Delta J$ into the wave, while the wave does work $\Omega(r)\Delta J$ on the disk when it deposits this amount of angular momentum into the fluid. The remaining energy is availabel as heat.

According to this equation, only irreversible heating contributes to the angular momentum exchange with the disk. This is easy to understand by noticing that after the fluid element has passed through the shock its angular momentum does not remain constant but {\it keeps being changed} by the post-shock azimuthal pressure gradient, arising because of the 2D post-shock fluid motions (ultimately resulting from the periodicity of the disk motion). This is the reason why we do not calculate the angular momentum transport by directly computing the change of the azimuthal velocity of the fluid at the shock from the jump conditions --- post-shock pressure gradients will wipe out most of this azimuthal velocity jump. Only after the fluid density returns to its unperturbed value $\rho_0$ (before passing through the shock again) can one measure the ultimate change of its angular momentum. And, as we discussed in \S \ref{sect:heating}, the only contribution to the change of internal energy of the fluid available at this point is the irreversible heating $\dot Q$. This fact has been used in \citet{GR01} and \citet{Belyaev2} to study the shock-driven disk evolution.

Combining equations (\ref{eq:heat_per_r}) and (\ref{eq:heat_J}) one finds the rate of angular momentum deposition by the shock into the disk fluid:
\ba
\frac{dF_J}{dr} &=& \mbox{sgn}[\Omega(r)-\Omega_P]~mr\Sigma c_0^2~\psi_Q(\Pi).
\label{eq:dFJdr}
\ea
Then it follows from equation (\ref{eq:Mdot}) that
\ba
\dot M(r) &=& \mbox{sgn}[\Omega(r)-\Omega_P]~\Omega\Sigma h^2
\nonumber\\
&\times & m\left(\frac{d\ln l}{d\ln r}\right)^{-1}\psi_Q(\Pi),
\label{eq:Mdot_fin}
\ea
where $h\equiv c_0/\Omega$ is the isothermal scaleheight of the disk. Our definition of $\dot M$ assumes that $\dot M>0$ for gas {\it inflow}. 

Equation (\ref{eq:Mdot_fin}) provides a desired relation between $\dot M$ and the local shock properties, namely the pressure ratio across the shock $\Pi$. In the weak shock limit (see \S \ref{sect:weak}) it reduces to the result of \citet{Belyaev2}. Note that $\dot M(r)$ depends only on the sign of $\Omega(r)-\Omega_P$ but not on its magnitude. Similar to $d\dot Q/dS$ it is insensitive to the wave pitch angle. We will use equation (\ref{eq:Mdot_fin}) to assess the shock-driven evolution of the disk in \S \ref{sect:accr_shock} \& \ref{sect:role_evol}.


\section{Discussion.}  
\label{sect:disc}


Results of the previous sections provide a framework for understanding the effects of global density waves on the protoplanetary disk properties and evolution. Our calculations are fully general: they are based on conservation of energy and momentum only (including the shock jump conditions) and do not depend on the exact nature of the perturber driving the spiral arms in the disk. 

Our results are formulated in terms of the pressure ratio at the shock $\Pi$, which is a function of the density ratio $\rho/\rho_0$, see equation (\ref{eq:Trho}). However, from observations one can hope to infer only the {\it surface density} perturbation at the spiral arm, not the variation of the 3D gas density. At the same time, right at the shock front $\Delta\rho/\rho_0$ should be equal to the surface density jump $\Delta\Sigma/\Sigma_0$. This is because immediately behind the shock vertical motions driven by the increase of gas pressure do not yet have a chance to puff the disk up and reduce $\Delta\rho/\rho_0$ compared to $\Delta\Sigma/\Sigma_0$. For that reason we will subsequently set $\Delta\rho/\rho_0=\Delta\Sigma/\Sigma_0$ and use equation (\ref{eq:Trho}) to express
\ba
\Pi=\frac{2+(\gamma+1)(\Delta\Sigma/\Sigma_0)}{2-(\gamma-1)(\Delta\Sigma/\Sigma_0)}.
\label{eq:PiSig}
\ea
We would then expect, for example, $\psi_Q\approx 0.1$ for $\Delta\Sigma/\Sigma_0=1$, see the estimates after equation (\ref{eq:Mach}). 

Determination of even $\Delta\Sigma/\Sigma$ is not an easy task. Previously \citet{Muto} and \citet{Grady} used surface brightness contrast of the spiral arms to infer the mass of the perturber, assuming the former to linearly scale with the latter and with $\Delta\Sigma/\Sigma_0$. However, the scattered light intensity is determined by the shape of the disk surface, which is more sensitive to the vertical gas motions than to the total disk surface density. Given the complexity and nonlinearity of the shock induced vertical motion we see little reason to expect $\Delta\Sigma/\Sigma_0$ to be equal to, or even to scale linearly with, the contrast of the scattered light surface brightness of the spiral arms \citep{Juhasz}. Because of that we urge caution in interpreting brightness variations across spirals in the images of the protoplanetary disks. In this work we  prefer to consider $\Delta\Sigma/\Sigma_0$ as a free parameter, although we do expect $\Delta\Sigma/\Sigma_0\sim 1$ for observed spirals.

A better way of determining $\Delta\Sigma/\Sigma_0$ may be to measure the gradient of the sub-millimeter emission across the spiral arm. Disk is expected to be optically thin at these wavelength and such emission should directly probe its surface density distribution \citep{ALMA_spirals}.

We now consider the implications of our findings for the thermodynamic state of the protoplanetary disks, their accretion rate, and timescale on which their surface density evolves, focusing on observational constraints. We also discuss the possibility of measuring the disk torque caused by the spiral arm self-gravity directly from observations and the implications for other types of accretion disks.


\subsection{Role of shock heating in thermal balance of the disk}  
\label{sect:role_heat}


Protoplanetary disks receive their energy from two main sources: irradiation by the central star and viscous dissipation. The latter is usually subdominant in protoplanetary disks, especially in their outer parts. However, in other types of astrophysical disks viscous heating dominate and we consider such systems in \S \ref{sect:other}. 

Dissipation at spiral shocks provides additional heating mechanism and here we estimate its role in determining the thermodynamical state of the disk. Focusing on protoplanetary disks for now, we compare the specific heating rate $d\dot Q/dS$ due to shocks, given by equation (\ref{eq:heat_per_S}) with the amount of energy $d\dot E_{\rm irr}/dS$ that a surface element of the disk receives from the central star. The latter is given by
\ba
\frac{d\dot E_{\rm irr}}{dS}=\zeta(r)\frac{L_\star}{4\pi r^2},
\label{eq:}
\ea
where $L_\star$ is the stellar luminosity and $\zeta$ is the angle at which the disk surface is illuminated by the radially-propagating starlight. For a standard passively irradiated disk \citep{chiang_1997} one has $\zeta(r) \approx (2\eta/7) (h/r)$, where parameter $\eta\sim 1-3$ characterizes the heght of the disk photosphere (in units of the disk scaleheight $h$). 

We can now assess the significance of the shock heating by forming a ratio
\ba
\frac{d\dot Q/dS}{d\dot E_{\rm irr}/dS} &=& 2m\zeta^{-1}\frac{|\Omega_P-\Omega(r)|\Sigma r^2 c_0^2}{L_\star}\psi_Q(\Pi).
\label{eq:heat_ratio}
\ea
For a passive, externally irradiated disk (this assumption implies that shock heating is a subdominant energy source, which we verify below) one finds
\ba
\frac{d\dot Q/dS}{d\dot E_{\rm irr}/dS} & \approx &\frac{7}{\eta}m\frac{\Sigma r^4 \Omega^3}{L_\star}\frac{h}{r}\left|1-\frac{\Omega_P}{\Omega}\right|\psi_Q(\Pi)
\label{eq:heat_ratio_passive}\\
&\approx & 10^{-2}~\frac{m}{2}\left|1-\frac{\Omega_P}{\Omega}\right|\frac{\psi_Q(\Pi)}{0.1}\frac{h/r}{0.1}
\nonumber\\
&\times &\frac{M_{\star,1}^{3/2}}{L_{\star,1}\eta_2}r_{50}^{-2},
\label{eq:est1}
\ea
where $M_{\star,1}\equiv M_\star/M_\odot$ is the scaled mass of the central star, $L_{\star,1}\equiv L_\star/L_\odot$, $\eta\equiv\eta/2$, and the numerical estimate assumes surface density profile (\ref{eq:MMSN}). If the spiral is produced by an external perturber orbiting at larger distance then $\Omega_P/\Omega(r)\ll 1$.

This estimate shows that shock heating is unlikely to provide a significant contribution to the thermal balance of the disk on scales of tens of AU that are probed by the direct imaging observations of protoplanetary disks in scattered light. Moreover, in this estimate we assumed rather high value of $\psi_Q=0.1$, which corresponds to the density jump $\Delta\Sigma/\Sigma_0\approx 1$ for $\gamma=7/5$, see
equation (\ref{eq:psiQ_rho}). Weaker shocks with smaller values of $\Delta \Sigma/\Sigma_0$ should produce negligible irreversible heat release compared to the stellar irradiation because of the steep dependence of $\psi_Q$ on density contrast in the limit $\Delta\Sigma/\Sigma_0=\Delta \rho/\rho_0\lesssim 1$, see \S \ref{sect:weak}. As a result, shock heating can likely be neglected in calculations of the protoplanetary disk structure at distances of tens of AU.

Situation may be different closer the terrestrial zone, at $r\lesssim 5$ AU. However, even there a reasonably strong shock is needed to produce $\Delta\Sigma/\Sigma_0\gtrsim 1$, leading to $\psi_Q\gtrsim 0.1$. Generally, one may not expect $\Delta\Sigma/\Sigma_0\gtrsim 1$ over a significant radial extent of the disk because shock damping tends to reduce $\Delta\Sigma/\Sigma_0$ to more moderate values.

Our finding of inefficiency of shock heating at warming the disk appears to be at odds with the recent simulations by \citet{Lyra1} and \citet{Lyra2}. These authors have found significant shock heating in their 2D and 3D simulations of disks with embedded massive ($5M_J$) planets. We believe this difference to be caused by the way in which shock heating was characterized in these works --- it was linked to the temperature jump behind the shock, which, as we showed here, measures the total internal energy release at the shock $\Delta U$ rather than the irreversible heat production $Q$, which is typically much smaller. 

We suggest that shock heating in simulations should be measured via the {\it entropy jump} at the shock $\Delta S$. Indeed, equation (\ref{eq:TsdS}) demonstrates that irreversible energy release at the shock can be written as
\ba
\frac{dQ}{dm}=\frac{k_BT_0}{(\gamma-1)\mu}\left[\exp\left(\Delta S/c_V\right)-1\right].
\label{eq:viaS}
\ea
Thus, measurements of the pre-shock gas temperature $T_0$ and $\Delta S$ at the shock\footnote{It is important to separate the entropy jump at the shock from other possible sources of entropy generation in simulations, e.g. numerical or artificial viscosity.} give a simple way of computing the irreversible shock heating.


\subsection{Shock-driven accretion}  
\label{sect:accr_shock}


Now we evaluate the efficiency of spiral shocks at driving mass accretion in the disk. To that effect we compare the shock induced accretion rate given by equation (\ref{eq:Mdot_fin}) with the mass accretion rate $\dot M_v$ due to viscous stresses. The latter is given by equation (\ref{eq:Mdot}) in which we need to replace $F_J$ with the {\it viscous angular momentum flux} $F_v\equiv 3\pi\nu\Sigma l$, where $\nu$ is the viscosity and the Keplerian rotation profile is assumed \citep{lynden-bell_1974,Rafikov13}. Then for the arbitrary radial distribution of $F_v(r)$ viscous accretion rate is
\ba
\dot M_v=3\pi\nu\Sigma\left(\frac{d\ln F_v}{d\ln r}\right)\left(\frac{d\ln l}{d\ln r}\right)^{-1}.
\label{eq:dotMvisc}
\ea
In standard constant $\dot M_v$ disks one has $F_v=\dot M_v l$ with radially constant $\dot M_v=3\pi\nu\Sigma$ \citep{Pringle}. However, in general $F_v$ can be an arbitrary function of $r$ (or $l$) \citep{Rafikov13,Vartanyan}.

Adopting standard viscosity parametrization $\nu=\alpha c_0^2/\Omega$ \citep{SS,Pringle} and using equations (\ref{eq:Mdot_fin}) and (\ref{eq:dotMvisc}) we can write
\ba
\frac{\dot M}{\dot M_v}=\frac{m}{3\pi}\frac{\psi_Q(\Pi)}{\alpha}\left(\frac{d\ln F_v}{d\ln r}\right)^{-1},
\label{eq:Mddot_rat}
\ea
where $|d\ln F_v/d\ln r|\sim 1$. This formula demonstrates that in a disk with the radially-constant viscous accretion rate $\dot M_v$ (so that $d\ln F_v/d\ln r=1/2$) shock-driven accretion rate dominates over the viscous one as long as $\psi(\Pi)\gtrsim 4.7(\alpha/m)$ locally. Thus, for $\alpha\lesssim 10^{-2}$ typically adopted for protoplanetary disks, spiral shocks with $\Delta\Sigma/\Sigma_0\sim 1$ would certainly dominate mass accretion through the disk compared to viscous stresses.

According to equation (\ref{eq:Mdot_fin}) the observed spiral shocks should give rise to mass accretion at the local rate 
\ba
|\dot M(r)|\approx 2\times 10^{-7}M_\odot~\mbox{yr}~\frac{m}{2}\left(\frac{h/r}{0.1}\right)^2\frac{\psi_Q(\Pi)}{0.1}r_{50}^{-1},
\label{eq:dotMest}
\ea
where we assumed surface density in the form (\ref{eq:MMSN}) and the  Keplerian profile of $\Omega(r)$. This estimate reveals quite rapid mass redistribution in disks with global spirals. At separations of $\sim 10$ AU this $\dot M$ exceeds typical accretion rates of T Tauri stars inferred from the UV/X-ray signatures \citep{Gullbring}. 

Of course, it has to be remembered that equation (\ref{eq:dotMest}) characterizes $\dot M$ on scales of tens of AU, and accretion rate onto the stellar surface may be quite different. Nevertheless, it is still interesting to compare theoretical $\dot M$ computed for observed spiral-bearing disks via equation (\ref{eq:dotMest}) to the actual rates of accretion onto the stellar surface, which we do next.

Disk around HD 100453 exhibits two spiral arms \citep{Wagner} at $r\sim 30$ AU, which are likely excited by the nearby stellar companion \citep{Dong_spiral}. Central star shows signs of weak accretion: \citet{Garcia} find $\dot M\sim 10^{-8}M_\odot$ yr$^{-1}$, while \citet{Collins} argue that $\dot M\lesssim 10^{-9}M_\odot$ yr$^{-1}$. For the low mass of its disk ($\lesssim 3\times 10^{-4}M_\odot$, \citet{Collins}) typical for the age of the system ($\sim 10$ Myr, \citet{Kama}) equation (\ref{eq:dotMest}) would predict $\dot M\approx 2\times 10^{-9}M_\odot$ yr$^{-1}$ at the location of the spirals, roughly consistent with observational constraints (keeping in mind strong dependence of $\psi_Q$ on rather uncertain $\Delta\Sigma/\Sigma_0$). 

MWC 758 (HD 36112) has a low mass disk ($\sim 3\times 10^{-3}M_\odot$, \citet{Chapillon}) with two prominent spiral arms at $r\sim 100$ AU \citep{Benisty}, and features high $\dot M\approx 10^{-6}M_\odot$ yr$^{-1}$ onto the star \citep{Donehew}. This rate exceeds shock-driven $\dot M$ computed via equation (\ref{eq:dotMest}) at $r=100$ AU by more than an order of magnitude.

SAO 206462 (HD 135344B) has a disk with $\Sigma\sim 20$ g cm$^{-1}$ at $r=50$ AU, where the spiral arms are located \citep{Garufi}. In this case shock-driven accretion predicts $\dot M\sim 4\times 10^{-7}M_\odot$ yr$^{-1}$, more than an order of magnitude higher than the observed rate of stellar accretion $\dot M\approx (1-4)\times 10^{-8}M_\odot$ yr$^{-1}$ \citep{Sitko}. 

Finally, low mass disk ($\sim 10^{-3}M_\odot$) around HD 100546 shows spirals at $\approx 200$ AU. Equation (\ref{eq:dotMest}) would then predict $\dot M\lesssim 10^{-8}M_\odot$ yr$^{-1}$, much lower than the observed $\dot M\approx  6\times 10^{-8}M_\odot$ yr$^{-1}$ \citep{Pogodin}. 

The discrepancy between the shock-driven $\dot M$ and the observed stellar accretion rate can have a number of causes. First, system parameters may be inaccurately determined. Second, $\Delta\Sigma/\Sigma_0$ can easily be different from unity, which is assumed in equation (\ref{eq:dotMest}). Given the strong sensitivity of $\psi_Q$ to the density jump at the shock (see Figures \ref{fig:7_5}b and \ref{fig:5_3}b) this may strongly change our estimate of the shock-driven $\dot M$. Third, in cases when the theoretical $\dot M$ is lower than the observed rate, there could be some additional angular momentum transport mechanism causing fast accretion in the inner disk. Finally, the two accretion rates can easily be different because any change of $\dot M$ propagates from the radii of tens of AU to the inner disk with some delay, which we calculate next.


\subsection{Shock-driven evolution of the disk}  
\label{sect:role_evol}


We can also use the results of \S \ref{sect:evolution} to study global disk evolution. It can be explored by combining equation (\ref{eq:Mdot_fin}) with the continuity equation
\ba
\frac{\partial\Sigma}{\partial t}+\frac{1}{2\pi r}\frac{\partial\dot M(r)}{\partial r}=0,
\label{eq:cont}
\ea
where we have neglected the viscous accretion rate $\dot M_v$ compared to $\dot M$, motivated by the results of \S \ref{sect:accr_shock}. Solving equation (\ref{eq:cont}) one can  determine $\Sigma(r,t)$ provided that the radial dependence of the pressure ratio at the shock $\Pi$ is specified for the whole disk. In practice, it may be non-trivial to compute $\Pi(r)$, since such calculation requires the knowledge of the details of the density wave excitation and nonlinear damping at the shock, both of which depend on $\Sigma(r)$ and $T(r)$. 

For these reason, in this work we limit ourselves to providing a simple and rather general estimate of the disk evolution timescale, but do not calculate $\Sigma(r,t)$ in detail. We define the {\it accretion timescale} as $t_{\rm acc}\equiv |\partial\ln\Sigma/\partial t|^{-1}$. Expressing $\partial\Sigma/\partial t$ from equation (\ref{eq:cont}), in which we approximate $\partial\dot M/\partial r\sim \dot M(r)/r$, and using equation (\ref{eq:Mdot_fin}) we obtain  
\ba
t_{\rm acc} &\approx &\frac{\pi}{m}\Omega^{-1}[\psi_Q(\Pi)]^{-1}\left(\frac{r}{h}\right)^2
\label{eq:tacc}\\
&\approx & 10^5~\mbox{yr}~\frac{2}{m}\frac{0.1}{\psi_Q(\Pi)}\left(\frac{h/r}{0.1}\right)^{-2}r_{50}^{3/2}.
\label{eq:tacc_est}
\ea
Comparing this expression with the standard estimate of the viscous evolution timescale $\Omega^{-1}\alpha^{-1}(r/h)^2$ one can see that shock driven evolution of the disk can be characterized by the effective dimensionless "shock viscosity"
\ba
\alpha_{\rm sh}=\frac{m}{\pi}\psi_Q(\Pi)\approx 0.03~\frac{m}{2}\frac{\psi_Q(\Pi)}{0.1}.
\label{eq:alpha_sh}
\ea
A similar expression for the effective shock-induced $\alpha$ has been obtained by \citet{Larson} based on the energy dissipation at the shock. In the limit of weak shocks equation (\ref{eq:alpha_sh}) predicts $\alpha_{\rm sh}\propto (\Delta\Sigma/\Sigma_0)^3$ (see equation (\ref{eq:Qweak}) in agreement with the results of \citet{Larson_conf} and \citet{Spruit}. However, the concept of $\alpha_{\rm sh}$ should be used with caution since the shock-driven angular momentum transport is intrinsically {\it global}, while the $\alpha$ ansatz is usually invoked to characterize the {\it local} shear viscosity.


\subsection{Implications for disk dispersal and planet formation}  
\label{sect:impl}


Estimate (\ref{eq:tacc_est}) demonstrates that strong spiral shocks can drive rapid disk evolution on large scales: even at $r=100$ AU this formula predicts $t_{\rm acc}\approx 0.3$ Myr, short compared to the characteristic disk lifetime of 3-7 Myr inferred from observations \citep{Kraus}. This suggests that disks exhibiting global spiral density waves with $\Delta\Sigma/\Sigma_0\sim 1$ (for which $\psi_Q\sim 0.1$, see equation (\ref{eq:psiQ_rho})) may be in the final stages of their dispersal via the shock-induced accretion, irrespective of the wave origin (i.e. whether they are driven by massive planets, stellar companions, or gravitational instability). 

Short timescale of the shock-driven disk evolution may have interesting implications for the origin of the transitional disks featuring large cavities devoid of dust \citep{Espaillat,Owen}. Equations (\ref{eq:dotMest}) and (\ref{eq:tacc_est}) suggest that the shock-driven $\dot M(r)$ increases and the evolution timescale $t_{\rm acc}$ decreases towards the star, as long as $\psi_Q$ does not decrease dramatically at small radii. This implies that a strong global wave should be depleting the disk by accretion in the inside-out fashion, with the inner disk losing mass on shorter time at higher $\dot M$ than the regions outside of it, resulting in the formation of a central cavity. 

In this picture, at a given age of the spiral $t_s$ --- time since it first emerged in the disk (which should be much shorter than the age of the system) --- the cavity edge would correspond to the radius $r_c(t_s)$ such that $t_{\rm acc}(r_c)=t_s$. The mass accretion rate through the cavity onto the star would be $\sim \dot M(r_c)$. As $t_s$ increases, so does $r_c$ (the cavity expands), while $\dot M$ onto the star would go down (while remaining quite high), according to equation (\ref{eq:dotMest}). When $r_c$ becomes very large and the outer disk gets severely depleted by the shock-induced accretion, photoevaporation \citep{Alexander} will likely disperse the remaining gas. 

Density drop at the cavity edge would stop the inflow of large dust particles via the usual mechanism of the dust filtration \citep{Paardekooper,Rice}, depleting the inner disk of these grains and making it transparent in submillimeter continuum observations \citep{Espaillat}. The gas would continue flowing through the cavity toward the star due to persistent shock dissipation to maintain accretion at the current rate $\dot M(t_s)$.

Note that in this picture the agent exciting the spiral shock (massive planet or stellar companion) should be located {\it outside} the cavity edge, naturally allowing gas to flow through the cavity and accrete onto star, as observed. This is different from the standard planetary clearing scenario for the origin of transitional disks \citep{Zhu_trans,Sally}, in which the cavity edge is assumed to correspond to the outer edge of a gap cleared by a planet (or planets) residing {\it inside} the cavity. In the latter scenario planetary tidal barrier suppresses gas inflow from the outer disk, preventing gas accretion onto the star at observed rates. 

It would be optimistic to expect our scenario of the shock-induced evolution to explain the origin of {\it all} transition disks. First, only a subset of directly imaged transitional disks feature global spirals. This problem could be partly alleviated by the fact that density waves may not be easy to detect in scattered light in general \citep{Juhasz}. But, second, driving high-amplitude global spirals likely requires massive companions (at least several $M_J$) to reside at separations of $\sim 10^2$ AU. However, according to the results of recent direct imaging surveys \citep{Nielsen,Brandt} only a small fraction of systems could harbor such companions. 

Nevertheless, the fact that most of the disks with spirals are in the transitional category (MWC 758, SAO 206462, HD 100546) suggests that at least in some cases shock-induced cavity clearing may be at work, possibly explaining at least some transitional disks in the ``mm-bright'' subclass of \citet{Owen}. A careful assessment of this possibility would require a global study of the behavior of the shock characteristics (radial profile of $\Pi$ entering $\psi_Q$ in equation (\ref{eq:Mdot_fin})), self-consistently coupled to the evolution of the disk properties.

Rapid pace of the shock-induced disk evolution may also shed light on the origin of giant planets. If a planet-induced spiral wave of moderate strength is observed in a disk around a relatively mature star (older than several Myr), then the short $t_{\rm acc}$ would imply that the planet has attained most of its mass relatively recently (within a few $t_{\rm acc}$) compared to the disk lifetime and that shocks are also relatively young. This would argue in favor of that planet having formed via the core accretion, which is expected to produce planets rather late. An alternative model of gravitational instability is thought to produce planets early in the disk lifetime; their density waves would have destroyed the disk on a much shorter timescale than the inferred age of the central star.

Weaker density waves with smaller values of $\Delta\Sigma/\Sigma_0\lesssim 1$ are far less efficient drivers of the disk evolution. Estimates made in \S \ref{sect:heating} imply that a wave with somewhat lower  $\Delta\Sigma/\Sigma_0=0.3$ would give rise to angular momentum transport with low $\alpha_{\rm sh}\sim 1.5\times 10^{-3}m$, in agreement with \citet{Spruit} and \citet{Larson}. This is comparable to the value of anomalous $\alpha$ even in the weakly ionized protoplanetary disks, in which MRI operates in the non-ideal regime \citep{Fleming}. This would extend the timescale of the shock-induced evolution of the disk at 100 AU to $\sim 6$ Myr, making it comparable to or exceeding the observed disk lifetimes. Previously, \citet{GR01} and \citet{Rafikov02} considered protoplanetary disk accretion driven by low mass (several $M_\oplus$) planets and found much lower values of $\dot M$ and effective $\alpha$ than suggested by estimates (\ref{eq:dotMest}) and (\ref{eq:alpha_sh}).


\subsection{Gravitational torque.}  
\label{sect:grav_torque}


One other piece of information that can be deduced from the images  of spiral-bearing disks is the strength of the torque that a non-axisymmetric spiral structure exerts onto itself due to its own self-gravity. Such {\it gravitational torque} was previously considered by \citet{Larson84} and \citet{Gnedin} in the context of protostellar disks and galactic spiral arms. 

\citet{BT} give the following expression for the magnitude of the gravitational torque that the disk outside radius $r$ exerts on the part of the disk interior to that radius in the WKB limit:
\ba
|C_{\rm G}(r)|=\pi^2m\frac{Gr\delta\Sigma^2}{k^2}=\frac{\left(\pi\tan\zeta\right)^2}{m}Gr^3\Sigma^2\left(\frac{\delta\Sigma}{\Sigma}\right)^2,
\label{eq:CG}
\ea
where they assumed that the non-axisymmetric component of $\Sigma$ has the form
\ba
\Sigma_{\rm na}(r,\varphi)=\delta\Sigma(r)\cos[m\varphi+f(r)].
\label{eq:non-ax}
\ea 
Here $f(r)$ gives the 2D shape of the spiral arm and $\delta\Sigma$ describes variation of its amplitude with radius. Radial wavenumber $k$ is related to $f(r)$ via $k=df/dr$, and to the pitch angle of spiral $\zeta$ via $\tan \zeta=m/(kr)$. A more general expression for $C_{\rm G}$ that applies to open spirals has been derived by \citet{Gnedin}. It features a correction factor $[1+(\tan\zeta)^2]^{-3/2}$, which is close to unity for spirals observed in protoplanetary disks ($<40\%$ correction for $\zeta<30^\circ$). 

Even though the azimuthal profile of the surface density in spirals seen in protoplanetary disks is not as simple as assumed in equation (\ref{eq:non-ax}) we will still use the result (\ref{eq:CG}) to estimate the strength of the gravitational torque. We do this by calculating the time $t_{\rm G}$ it would take to remove the enclosed angular momentum of the disk $L(r)\equiv 2\pi\int_0^rr^\prime\Sigma(r^\prime)l(r^\prime)dr^\prime=2\pi r\Sigma l$ (for $\Sigma$ given by equation (\ref{eq:MMSN})) by the gravitational torque alone:
\ba
t_{\rm G}&\equiv &\frac{L(r)}{|C_{\rm G}(r)|}\approx \frac{2m}{\pi\tan^2\zeta}\Omega^{-1}\frac{M_\star}{\Sigma r^2}\left(\frac{\Delta\Sigma}{\Sigma_0}\right)^{-2}
\label{eq:tG}\\
&\approx & 2\times 10^5\mbox{yr}~\frac{m}{2}\left(\frac{\tan 20^\circ}{\tan\zeta}\right)^2\left(\frac{\Delta\Sigma}{\Sigma_0}\right)^{-2}r_{50},
\ea
where we took $\Sigma(r)$ from equation (\ref{eq:MMSN}). In this estimate we identified $\delta\Sigma$ with the surface density jump at the shock $\Delta\Sigma/\Sigma_0=\Delta\rho/\rho_0$. Note, that equations (\ref{eq:CG}) and (\ref{eq:tG}) do not require information about the thermodynamic properties of the disk and contain variables ($\Delta\Sigma$, $\Sigma_0$, $\zeta$), which are potentially directly measurable from observations.

It should not be surprising that $t_{\rm G}>t_{\rm acc}$ at $r\lesssim 100$ AU. First, $t_{\rm G}$ characterizes density wave {\it excitation}, while $t_{\rm acc}$ described angular momentum {\it deposition} into the disk fluid by the wave. These are separate processes and, strictly speaking, their rates should not be directly compared. Second, gravitational torque is only a part of the full excitation torque experienced by the spiral wave \citep{Larson_conf}. The rest is provided by the coupling of its non-axisymmetric density pattern with the potential of the perturber (and not the potential of the arm itself, as is the case for the gravitational torque). The relative contribution of $C_{\rm G}$ scales inversely with the Toomre Q \citep{Larson_conf}, which is pretty high, $Q_T\sim 10$, for the disk with $\Sigma$ given by equation (\ref{eq:MMSN}) and a star with $L_\star=L_\odot$ even at 100 AU. 

Nevertheless, a measurement of $C_{\rm G}$ can still give us a direct {\it upper limit} on the disk evolution timescale via the equation (\ref{eq:tG}), which is a useful piece of information from the evolutionary standpoint, see \S \ref{sect:impl}.


\subsection{Spiral shocks in other types of astrophysical disks}  
\label{sect:other}


Spiral structures are not unique to protoplanetary disks and have been found in other types of accretion disks. In particular, the technique of Doppler tomography \citep{Marsh} reveals spiral arms in disks of cataclysmic variables \citep{Spiral} as well as in the debris disk around the isolated metal-rich white dwarf SDSS J122859.93+104032.9 \citep{Manser}. Also, simulations of circumbinary gaseous disks orbiting the supermassive black hole (SMBH) binaries \citep{Milos} and stellar binaries \citep{Lines} routinely reveal spiral arms. Quite naturally, most of the results of this work can be directly applied to these systems, with one important exception. 

In \S \ref{sect:role_heat} we compared shock heating of the protoplanetary disks with their irradiation by the central star. However, irradiation is generally negligible for disks around compact objects (cataclysmic variables and X-ray binaries) and SMBH binaries, which are heated primarily by viscous heating. Thus, to assess the role of shock heating for these objects we need to compare $d\dot Q/dS$ with the rate of viscous heating per unit area $d\dot E_v/dS=(9/4)\alpha\Omega\Sigma c_0^2$, where we used $\alpha$-parametrization of viscosity \citep{Pringle}. Using equation (\ref{eq:heat_per_S}) we can write
\ba
\frac{d\dot Q/dS}{d\dot E_v/dS} & \approx &\frac{2m}{9\pi}\left|1-\frac{\Omega_P}{\Omega}\right|
\frac{\psi_Q(\Pi)}{\alpha}
\label{eq:heat_ratio_active}\\
&\approx & 1.4~\frac{m}{2}\left|1-\frac{\Omega_P}{\Omega}\right|
\frac{10^{-2}}{\alpha}\frac{\psi_Q(\Pi)}{0.1}.
\label{eq:est2}
\ea
This ratio is similar to the ratio of mass accretion rates in equation (\ref{eq:Mddot_rat}). This should not be surprising since both viscous heating and $\dot M_v$ owe their origin to the same process --- viscous stresses in the disk.

Assuming $\Omega\gg\Omega_P$, equation (\ref{eq:heat_ratio_active}) predicts that shock heating dominates over viscous dissipation as long as $\psi_Q(\Pi)\gtrsim 14(\alpha/m)$. However, in hot accretion disks this inequality presents a more stringent constraint on the shock strength because of higher values of $\alpha\sim 0.1-0.3$ expected in such disks \citep{Smak98,Smak99}. Nevertheless, even accounting for that it is clear that spiral waves with $\Delta\Sigma/\Sigma_0\sim 1$ should still be contributing at the level of (at least) tens of per cent to the thermal energy balance of the disk. Thus, shock heating should be taken into consideration in such disks, especially in systems possessing massive external perturbers capable of naturally driving strong spiral density waves (e.g. cataclysmic variables or SMBH binaries).

Irradiation is still important for the thermal balance of protoplanetary disks around stellar mass binaries, which are thought to be the birth sites of the circumbinary planets recently discovered by the {\it Kepler} mission \citep{Welsh}. However, on sub-AU scales, where one expects density waves driven by the central binary to operate, irradiation should be a subdominant heating agent compared to both the viscous dissipation and the tidal heating --- damping of the binary-driven density waves, as shown by \citet{Vartanyan}. This agrees with our  equation (\ref{eq:est1}), which predicts that shock heating should exceed central irradiation in stellar circumbinary disks for $r\lesssim 1$ AU.


\section{Summary.}  
\label{sect:summ}


We explored the effects of the global density waves on thermal state, accretion, and global evolution of protoplanetary disks. Such waves often stand out as spiral arms in high-resolution scattered light images of disks around young stars. Nonlinear effects accompanying propagation of these waves lead to their evolution into shocks, resulting in irreversible energy release and angular momentum transport across the disk. 

We developed analytical framework to characterize the heating, angular momentum deposition and mass accretion caused by the density waves as a function of their density (or pressure) contrast, applicable to shocks of arbitrary strength and origin. We demonstrate the importance of distinguishing between the total change of internal energy $\Delta U$ and the irreversible energy release at the shock $Q$ (\S \ref{sect:heating}). Only the latter provides the heat source for the disk and affects its thermodynamic balance. However, the former may still be affecting energy budget of the disk in indirect ways (\S \ref{sect:heat_rate}).

Irreversible heating in weak shocks with density contrasts $\Delta\Sigma/\Sigma_0\ll 1$ is negligible (scaling as $(\Delta\Sigma/\Sigma_0)^3$) compared to the jump of internal energy across the shock $\Delta U\propto \Delta\Sigma/\Sigma_0$ (\S \ref{sect:weak}). But even for strong shocks with pressure contrast $p/p_0\gg 1$ the former represents only a fraction (tens of per cent) of the latter (\S \ref{sect:strong}). 

Disk fluid passing through shocks experiences a change of its angular momentum, leading to accretion onto the star in the part of the disk where $\Omega$ is higher than the wave pattern speed $\Omega_P$. We analytically compute both the angular momentum deposition by the shock and the shock-induced accretion rate as a function of shock strength (\S \ref{sect:evolution}). 

Application of our results to observed protoplanetary disks leads to the following conclusions.

\begin{itemize}
    
    \item Shock-induced irreversible heating should be a minor contributor to the thermal balance of the passive disks irradiated by the central stars at separations of tens of AU, even for shocks of moderate strength, $\Delta\Sigma/\Sigma_0\sim 1$ (\S \ref{sect:role_heat}). Only within several AU from the star can the heating by shocks of moderate strength compete with stellar irradiation. For waves of lower amplitude shock heating should be negligible throughout the disk. 

\item At the same time, spiral waves drive mass accretion through the disk at rates, which can easily exceed the accretion rate due to viscous stresses (\S \ref{sect:accr_shock}). For MMSN-like surface density profile the wave-driven $\dot M$ can reach $10^{-7}M_\odot$ yr$^{-1}$ even at 100 AU. 

\item Shock-induced angular momentum transport can play important role for the global evolution of the disk (\S \ref{sect:role_evol}). For standard disk properties ($h/r\sim 0.1$) spiral waves of moderate amplitude ($\Delta\Sigma/\Sigma_0\sim 1$) are capable of driving significant surface density evolution within 0.5 Myr even at separations of 100 AU (and $\sim 10^4$ yr at 10 AU). 

\item Prominent spiral arms detected in scattered light may be the signposts of a protoplanetary disk disruption by the embedded massive planets or stellar companions on a timescale short compared to the disk lifetime. The planet- (or spiral-)induced disk dispersal can provide an alternative mechanism for the disk removal, complementary to its photoevaporation by the stellar UV and X-rays. Shock-induced evolution may proceed in the inside-out fashion, by first clearing the central cavity and, thus, naturally explaining transitional disk morphology of many spiral-bearing systems (\S \ref{sect:impl}).

\item Self-gravity of the non-axysimmetric density perturbation of the spiral wave gives rise to the gravitational torque acting on the disk. Its measurement from the imaging data allows one to set an upper limit on the wave-induced disk evolution timescale (\S \ref{sect:grav_torque}).

\item Our results can be naturally extended to other types of accreting systems. In particular, we showed that shock heating can provide a significant (up to tens of per cent) contribution to the energy budget of disks in cataclysmic variables and X-ray binaries, which are heated mainly by viscous dissipation (\S \ref{sect:other}).

\end{itemize}

Our results provide a basis for understanding the effects of the global density waves of arbitrary amplitude on the underlying disk structure and on the evolution of the waves themselves. 

\acknowledgements

Author is grateful to Scott Tremaine for stimulating discussions and to Ruobing Dong for useful suggestions. R.R.R. is an IBM Einstein Fellow at the IAS. Financial support for this study has been provided by NSF via grants AST-1409524,  AST-1515763, NASA via grant 14-ATP14-0059, and The Ambrose Monell Foundation.


\bibliographystyle{apj}
\bibliography{mybib}

\begin{thebibliography}{}
\expandafter\ifx\csname natexlab\endcsname\relax\def\natexlab#1{#1}\fi

\bibitem[{{Alexander} {et~al.}(2014){Alexander}, {Pascucci}, {Andrews},
  {Armitage}, \& {Cieza}}]{Alexander}
{Alexander}, R., {Pascucci}, I., {Andrews}, S., {Armitage}, P., \& {Cieza}, L.
  2014, Protostars and Planets VI, 475

\bibitem[{{Belyaev} {et~al.}(2013){Belyaev}, {Rafikov}, \& {Stone}}]{Belyaev2}
{Belyaev}, M.~A., {Rafikov}, R.~R., \& {Stone}, J.~M. 2013, \apj, 770, 67

\bibitem[{{Benisty} {et~al.}(2015){Benisty}, {Juhasz}, {Boccaletti},
  {Avenhaus}, {Milli}, {Thalmann}, {Dominik}, {Pinilla}, {Buenzli}, {Pohl},
  {Beuzit}, {Birnstiel}, {de Boer}, {Bonnefoy}, {Chauvin}, {Christiaens},
  {Garufi}, {Grady}, {Henning}, {Huelamo}, {Isella}, {Langlois}, {M{\'e}nard},
  {Mouillet}, {Olofsson}, {Pantin}, {Pinte}, \& {Pueyo}}]{Benisty}
{Benisty}, M., {Juhasz}, A., {Boccaletti}, A., {et~al.} 2015, \aap, 578, L6

\bibitem[{{Binney} \& {Tremaine}(2008)}]{BT}
{Binney}, J., \& {Tremaine}, S. 2008, {Galactic Dynamics: Second Edition}
  (Princeton University Press)

\bibitem[{{Boccaletti} {et~al.}(2013){Boccaletti}, {Pantin}, {Lagrange},
  {Augereau}, {Meheut}, \& {Quanz}}]{Boccaletti}
{Boccaletti}, A., {Pantin}, E., {Lagrange}, A.-M., {et~al.} 2013, \aap, 560,
  A20

\bibitem[{{Boley} \& {Durisen}(2006)}]{Boley}
{Boley}, A.~C., \& {Durisen}, R.~H. 2006, \apj, 641, 534

\bibitem[{{Brandt} {et~al.}(2014){Brandt}, {McElwain}, {Turner}, {Mede},
  {Spiegel}, {Kuzuhara}, {Schlieder}, {Wisniewski}, {Abe}, {Biller},
  {Brandner}, {Carson}, {Currie}, {Egner}, {Feldt}, {Golota}, {Goto}, {Grady},
  {Guyon}, {Hashimoto}, {Hayano}, {Hayashi}, {Hayashi}, {Henning}, {Hodapp},
  {Inutsuka}, {Ishii}, {Iye}, {Janson}, {Kandori}, {Knapp}, {Kudo}, {Kusakabe},
  {Kwon}, {Matsuo}, {Miyama}, {Morino}, {Moro-Mart{\'{\i}}n}, {Nishimura},
  {Pyo}, {Serabyn}, {Suto}, {Suzuki}, {Takami}, {Takato}, {Terada}, {Thalmann},
  {Tomono}, {Watanabe}, {Yamada}, {Takami}, {Usuda}, \& {Tamura}}]{Brandt}
{Brandt}, T.~D., {McElwain}, M.~W., {Turner}, E.~L., {et~al.} 2014, \apj, 794,
  159

\bibitem[{{Chapillon} {et~al.}(2008){Chapillon}, {Guilloteau}, {Dutrey}, \&
  {Pi{\'e}tu}}]{Chapillon}
{Chapillon}, E., {Guilloteau}, S., {Dutrey}, A., \& {Pi{\'e}tu}, V. 2008, \aap,
  488, 565

\bibitem[{{Chiang} \& {Goldreich}(1997)}]{chiang_1997}
{Chiang}, E.~I., \& {Goldreich}, P. 1997, \apj, 490, 368

\bibitem[{{Christiaens} {et~al.}(2014){Christiaens}, {Casassus}, {Perez}, {van
  der Plas}, \& {M{\'e}nard}}]{ALMA_spirals}
{Christiaens}, V., {Casassus}, S., {Perez}, S., {van der Plas}, G., \&
  {M{\'e}nard}, F. 2014, \apjl, 785, L12

\bibitem[{{Collins} {et~al.}(2009){Collins}, {Grady}, {Hamaguchi},
  {Wisniewski}, {Brittain}, {Sitko}, {Carpenter}, {Williams}, {Mathews},
  {Williger}, {van Boekel}, {Carmona}, {Henning}, {van den Ancker}, {Meeus},
  {Chen}, {Petre}, \& {Woodgate}}]{Collins}
{Collins}, K.~A., {Grady}, C.~A., {Hamaguchi}, K., {et~al.} 2009, \apj, 697,
  557

\bibitem[{{Dodson-Robinson} \& {Salyk}(2011)}]{Sally}
{Dodson-Robinson}, S.~E., \& {Salyk}, C. 2011, \apj, 738, 131

\bibitem[{{Donehew} \& {Brittain}(2011)}]{Donehew}
{Donehew}, B., \& {Brittain}, S. 2011, \aj, 141, 46

\bibitem[{{Dong} {et~al.}(2015{\natexlab{a}}){Dong}, {Zhu}, {Fung}, {Rafikov},
  {Chiang}, \& {Wagner}}]{Dong_star}
{Dong}, R., {Zhu}, Z., {Fung}, J., {et~al.} 2015{\natexlab{a}}, ArXiv e-prints,
  arXiv:1512.04949

\bibitem[{{Dong} {et~al.}(2015{\natexlab{b}}){Dong}, {Zhu}, {Rafikov}, \&
  {Stone}}]{Dong_spiral}
{Dong}, R., {Zhu}, Z., {Rafikov}, R.~R., \& {Stone}, J.~M. 2015{\natexlab{b}},
  \apjl, 809, L5

\bibitem[{{Espaillat} {et~al.}(2014){Espaillat}, {Muzerolle}, {Najita},
  {Andrews}, {Zhu}, {Calvet}, {Kraus}, {Hashimoto}, {Kraus}, \&
  {D'Alessio}}]{Espaillat}
{Espaillat}, C., {Muzerolle}, J., {Najita}, J., {et~al.} 2014, Protostars and
  Planets VI, 497

\bibitem[{{Fleming} \& {Stone}(2003)}]{Fleming}
{Fleming}, T., \& {Stone}, J.~M. 2003, \apj, 585, 908

\bibitem[{{Garcia Lopez} {et~al.}(2006){Garcia Lopez}, {Natta}, {Testi}, \&
  {Habart}}]{Garcia}
{Garcia Lopez}, R., {Natta}, A., {Testi}, L., \& {Habart}, E. 2006, \aap, 459,
  837

\bibitem[{{Garufi} {et~al.}(2013){Garufi}, {Quanz}, {Avenhaus}, {Buenzli},
  {Dominik}, {Meru}, {Meyer}, {Pinilla}, {Schmid}, \& {Wolf}}]{Garufi}
{Garufi}, A., {Quanz}, S.~P., {Avenhaus}, H., {et~al.} 2013, \aap, 560, A105

\bibitem[{{Gnedin} {et~al.}(1995){Gnedin}, {Goodman}, \& {Frei}}]{Gnedin}
{Gnedin}, O.~Y., {Goodman}, J., \& {Frei}, Z. 1995, \aj, 110, 1105

\bibitem[{{Goodman} \& {Rafikov}(2001)}]{GR01}
{Goodman}, J., \& {Rafikov}, R.~R. 2001, \apj, 552, 793

\bibitem[{{Grady} {et~al.}(2013){Grady}, {Muto}, {Hashimoto}, {Fukagawa},
  {Currie}, {Biller}, {Thalmann}, {Sitko}, {Russell}, {Wisniewski}, {Dong},
  {Kwon}, {Sai}, {Hornbeck}, {Schneider}, {Hines}, {Moro Mart{\'{\i}}n},
  {Feldt}, {Henning}, {Pott}, {Bonnefoy}, {Bouwman}, {Lacour}, {Mueller},
  {Juh{\'a}sz}, {Crida}, {Chauvin}, {Andrews}, {Wilner}, {Kraus}, {Dahm},
  {Robitaille}, {Jang-Condell}, {Abe}, {Akiyama}, {Brandner}, {Brandt},
  {Carson}, {Egner}, {Follette}, {Goto}, {Guyon}, {Hayano}, {Hayashi},
  {Hayashi}, {Hodapp}, {Ishii}, {Iye}, {Janson}, {Kandori}, {Knapp}, {Kudo},
  {Kusakabe}, {Kuzuhara}, {Mayama}, {McElwain}, {Matsuo}, {Miyama}, {Morino},
  {Nishimura}, {Pyo}, {Serabyn}, {Suto}, {Suzuki}, {Takami}, {Takato},
  {Terada}, {Tomono}, {Turner}, {Watanabe}, {Yamada}, {Takami}, {Usuda}, \&
  {Tamura}}]{Grady}
{Grady}, C.~A., {Muto}, T., {Hashimoto}, J., {et~al.} 2013, \apj, 762, 48

\bibitem[{{Gullbring} {et~al.}(1998){Gullbring}, {Hartmann}, {Brice{\~n}o}, \&
  {Calvet}}]{Gullbring}
{Gullbring}, E., {Hartmann}, L., {Brice{\~n}o}, C., \& {Calvet}, N. 1998, \apj,
  492, 323

\bibitem[{{Juh{\'a}sz} {et~al.}(2015){Juh{\'a}sz}, {Benisty}, {Pohl},
  {Dullemond}, {Dominik}, \& {Paardekooper}}]{Juhasz}
{Juh{\'a}sz}, A., {Benisty}, M., {Pohl}, A., {et~al.} 2015, \mnras, 451, 1147

\bibitem[{{Kama} {et~al.}(2015){Kama}, {Folsom}, \& {Pinilla}}]{Kama}
{Kama}, M., {Folsom}, C.~P., \& {Pinilla}, P. 2015, \aap, 582, L10

\bibitem[{{Kraus} {et~al.}(2012){Kraus}, {Ireland}, {Hillenbrand}, \&
  {Martinache}}]{Kraus}
{Kraus}, A.~L., {Ireland}, M.~J., {Hillenbrand}, L.~A., \& {Martinache}, F.
  2012, \apj, 745, 19

\bibitem[{{Landau} \& {Lifshitz}(1959)}]{LL}
{Landau}, L.~D., \& {Lifshitz}, E.~M. 1959, {Fluid mechanics}

\bibitem[{{Larson}(1984)}]{Larson84}
{Larson}, R.~B. 1984, \mnras, 206, 197

\bibitem[{{Larson}(1989)}]{Larson_conf}
{Larson}, R.~B. 1989, in The Formation and Evolution of Planetary Systems, ed.
  H.~A. {Weaver} \& L.~{Danly}, 31--48

\bibitem[{{Larson}(1990)}]{Larson}
---. 1990, \mnras, 243, 588

\bibitem[{{Lines} {et~al.}(2015){Lines}, {Leinhardt}, {Baruteau},
  {Paardekooper}, \& {Carter}}]{Lines}
{Lines}, S., {Leinhardt}, Z.~M., {Baruteau}, C., {Paardekooper}, S.-J., \&
  {Carter}, P.~J. 2015, \aap, 582, A5

\bibitem[{{Lynden-Bell} \& {Pringle}(1974)}]{lynden-bell_1974}
{Lynden-Bell}, D., \& {Pringle}, J.~E. 1974, \mnras, 168, 603

\bibitem[{{Lyra} {et~al.}(2015){Lyra}, {Richert}, {Boley}, {Turner}, {Mac Low},
  {Okuzumi}, \& {Flock}}]{Lyra2}
{Lyra}, W., {Richert}, A.~J.~W., {Boley}, A., {et~al.} 2015, ArXiv e-prints,
  arXiv:1511.02988

\bibitem[{{MacFadyen} \& {Milosavljevi{\'c}}(2008)}]{Milos}
{MacFadyen}, A.~I., \& {Milosavljevi{\'c}}, M. 2008, \apj, 672, 83

\bibitem[{{Manser} {et~al.}(2016){Manser}, {G{\"a}nsicke}, {Marsh}, {Veras},
  {Koester}, {Breedt}, {Pala}, {Parsons}, \& {Southworth}}]{Manser}
{Manser}, C.~J., {G{\"a}nsicke}, B.~T., {Marsh}, T.~R., {et~al.} 2016, \mnras,
  455, 4467

\bibitem[{{Marino} {et~al.}(2015){Marino}, {Perez}, \& {Casassus}}]{Marino}
{Marino}, S., {Perez}, S., \& {Casassus}, S. 2015, \apjl, 798, L44

\bibitem[{{Marsh} \& {Horne}(1988)}]{Marsh}
{Marsh}, T.~R., \& {Horne}, K. 1988, \mnras, 235, 269

\bibitem[{{Muto} {et~al.}(2012){Muto}, {Grady}, {Hashimoto}, {Fukagawa},
  {Hornbeck}, {Sitko}, {Russell}, {Werren}, {Cur{\'e}}, {Currie}, {Ohashi},
  {Okamoto}, {Momose}, {Honda}, {Inutsuka}, {Takeuchi}, {Dong}, {Abe},
  {Brandner}, {Brandt}, {Carson}, {Egner}, {Feldt}, {Fukue}, {Goto}, {Guyon},
  {Hayano}, {Hayashi}, {Hayashi}, {Henning}, {Hodapp}, {Ishii}, {Iye},
  {Janson}, {Kandori}, {Knapp}, {Kudo}, {Kusakabe}, {Kuzuhara}, {Matsuo},
  {Mayama}, {McElwain}, {Miyama}, {Morino}, {Moro-Martin}, {Nishimura}, {Pyo},
  {Serabyn}, {Suto}, {Suzuki}, {Takami}, {Takato}, {Terada}, {Thalmann},
  {Tomono}, {Turner}, {Watanabe}, {Wisniewski}, {Yamada}, {Takami}, {Usuda}, \&
  {Tamura}}]{Muto}
{Muto}, T., {Grady}, C.~A., {Hashimoto}, J., {et~al.} 2012, \apjl, 748, L22

\bibitem[{{Nielsen} {et~al.}(2013){Nielsen}, {Liu}, {Wahhaj}, {Biller},
  {Hayward}, {Close}, {Males}, {Skemer}, {Chun}, {Ftaclas}, {Alencar},
  {Artymowicz}, {Boss}, {Clarke}, {de Gouveia Dal Pino}, {Gregorio-Hetem},
  {Hartung}, {Ida}, {Kuchner}, {Lin}, {Reid}, {Shkolnik}, {Tecza}, {Thatte}, \&
  {Toomey}}]{Nielsen}
{Nielsen}, E.~L., {Liu}, M.~C., {Wahhaj}, Z., {et~al.} 2013, \apj, 776, 4

\bibitem[{{Owen}(2015)}]{Owen}
{Owen}, J.~E. 2015, ArXiv e-prints, arXiv:1512.06873

\bibitem[{{Paardekooper} \& {Mellema}(2004)}]{Paardekooper}
{Paardekooper}, S.-J., \& {Mellema}, G. 2004, \aap, 425, L9

\bibitem[{{Pogodin} {et~al.}(2012){Pogodin}, {Hubrig}, {Yudin}, {Sch{\"o}ller},
  {Gonz{\'a}lez}, \& {Stelzer}}]{Pogodin}
{Pogodin}, M.~A., {Hubrig}, S., {Yudin}, R.~V., {et~al.} 2012, Astronomische
  Nachrichten, 333, 594

\bibitem[{{Pringle}(1981)}]{Pringle}
{Pringle}, J.~E. 1981, \araa, 19, 137

\bibitem[{{Rafikov}(2002)}]{Rafikov02}
{Rafikov}, R.~R. 2002, \apj, 569, 997

\bibitem[{{Rafikov}(2013)}]{Rafikov13}
---. 2013, \apj, 774, 144

\bibitem[{{Rice} {et~al.}(2006){Rice}, {Armitage}, {Wood}, \& {Lodato}}]{Rice}
{Rice}, W.~K.~M., {Armitage}, P.~J., {Wood}, K., \& {Lodato}, G. 2006, \mnras,
  373, 1619

\bibitem[{{Richert} {et~al.}(2015){Richert}, {Lyra}, {Boley}, {Mac Low}, \&
  {Turner}}]{Lyra1}
{Richert}, A.~J.~W., {Lyra}, W., {Boley}, A., {Mac Low}, M.-M., \& {Turner}, N.
  2015, \apj, 804, 95

\bibitem[{{Shakura} \& {Sunyaev}(1973)}]{SS}
{Shakura}, N.~I., \& {Sunyaev}, R.~A. 1973, \aap, 24, 337

\bibitem[{{Sitko} {et~al.}(2012){Sitko}, {Day}, {Kimes}, {Beerman}, {Martus},
  {Lynch}, {Russell}, {Grady}, {Schneider}, {Lisse}, {Nuth}, {Cur{\'e}},
  {Henden}, {Kraus}, {Motta}, {Tamura}, {Hornbeck}, {Williger}, \&
  {Fugazza}}]{Sitko}
{Sitko}, M.~L., {Day}, A.~N., {Kimes}, R.~L., {et~al.} 2012, \apj, 745, 29

\bibitem[{{Smak}(1999)}]{Smak99}
{Smak}, J. 1999, \actaa, 49, 391

\bibitem[{{Smak}(1998)}]{Smak98}
{Smak}, J.~I. 1998, \actaa, 48, 677

\bibitem[{{Spruit}(1987)}]{Spruit}
{Spruit}, H.~C. 1987, \aap, 184, 173

\bibitem[{{Steeghs} {et~al.}(1997){Steeghs}, {Harlaftis}, \& {Horne}}]{Spiral}
{Steeghs}, D., {Harlaftis}, E.~T., \& {Horne}, K. 1997, \mnras, 290, L28

\bibitem[{{Vartanyan} {et~al.}(2015){Vartanyan}, {Garmilla}, \&
  {Rafikov}}]{Vartanyan}
{Vartanyan}, D., {Garmilla}, J.~A., \& {Rafikov}, R.~R. 2015, ArXiv e-prints,
  arXiv:1509.07524

\bibitem[{{Wagner} {et~al.}(2015){Wagner}, {Apai}, {Kasper}, \&
  {Robberto}}]{Wagner}
{Wagner}, K., {Apai}, D., {Kasper}, M., \& {Robberto}, M. 2015, \apjl, 813, L2

\bibitem[{{Welsh} {et~al.}(2014){Welsh}, {Orosz}, {Carter}, \&
  {Fabrycky}}]{Welsh}
{Welsh}, W.~F., {Orosz}, J.~A., {Carter}, J.~A., \& {Fabrycky}, D.~C. 2014, in
  IAU Symposium, Vol. 293, IAU Symposium, ed. N.~{Haghighipour}, 125--132

\bibitem[{{Zhu} {et~al.}(2015){Zhu}, {Dong}, {Stone}, \& {Rafikov}}]{Zhu}
{Zhu}, Z., {Dong}, R., {Stone}, J.~M., \& {Rafikov}, R.~R. 2015, \apj, 813, 88

\bibitem[{{Zhu} {et~al.}(2011){Zhu}, {Nelson}, {Hartmann}, {Espaillat}, \&
  {Calvet}}]{Zhu_trans}
{Zhu}, Z., {Nelson}, R.~P., {Hartmann}, L., {Espaillat}, C., \& {Calvet}, N.
  2011, \apj, 729, 47

\end{thebibliography}

\end{document}